# Synthesis of FeNi tetrataenite phase by means of chemical precipitation

Kurichenko Vladislav L.[1], Karpenkov Dmitriy Yu.[1,2], Karpenkov Alexey Yu.[2,3], Lyakhova Marina B.[2,3], Khovaylo Vladimir V.[1]

[1] National University of Science and Technology "MISIS", Leninskiy prospect, 4, 119991 Moscow, Russia
[2] Chelyabinsk State University, Bratiev Kashirinykh street, 129, 454001 Chelyabinsk, Russia
[3] Tver State University, Zhelyabova Str., 33, 170100, Tver, Russia

E-mail: vkurichenko@misis.ru

**Abstract**

FeNi L10 (tetrataenite) phase has great perspectives for hard magnetic materials production. In this paper we report on synthesis of this phase in chemically co-precipitated FeNi nanopowder by means of a thermal treatment procedure which includes cycling oxidation and reduction processes at 320 °C. The presence of the FeNi L10 phase in the samples was confirmed by magnetic measurements and differential scanning calorimetry analysis.

## 1. Introduction

Due to the so-called "rare-earth crisis", much effort has been made recently in order to produce new high-energy permanent magnets. As a consequence of China' decision to restrict considerably the sale of rare earth elements, the price of the most technologically important $Nd_{12}Fe_{14}B$ magnets sharply increased. This has forced scientists to search new approaches for stabilization of already known hard magnetic phases as well as to develop novel high-performance magnetic materials. Table 1 illustrates properties of the known rare earth free magnetic systems. It is seen that among all the systems, FeNi $L1_0$ (tetrataenite) phase has the highest theoretical value of energy product (320 kJ/m$^3$), which is the figure of merit of a hard magnetic material. Thus, this phase has been considered as the most promising substitution material for permanent magnets application. However, the search for substitutive materials requires development of new approaches for extreme magnetic properties formation. Thus, the primarily aim of our work is to explore some new techniques for production of the desired phase.

Since the discovery of the FeNi $L1_0$ phase in the 1960s [5], various methods have been utilized to obtain this thermodynamically stable structure. Specifically, attempts to produce the tetrataenite phase by thin film deposition [6], high-pressure torsion [7], crystallization of as-quenched $Fe_{42}Ni_{41.3}Si_8B_4P_4Cu_{0.7}$ alloy [8], electroless nickel plating of Fe with subsequent cyclic heat treatment [9] have been reported in the literature. The main difficulty of the synthesis of this phase is extremely slow diffusion rate at temperatures below order-disorder transition temperature $T_{eq}$ = 320 °C. Above this temperature, the disordered face centered cubic A1-type structure is entropically stabilized. This is why the tetrataenite has been observed in nature only in meteorites, as they had plenty of time for long-range diffusion to happen. Therefore, the main goal of laboratory synthesis of tetrataenite is to increase diffusion rate at temperatures where this phase is stable.

**Table 1.** Magnetic properties of known rare-earth free phases.

| Phase | Maximum energy product $(BH)_{max}$, kJ/m$^3$ | Curie temperature, $T_c$, K | Saturation polarization $M_s$, MA/m | Uniaxial anisotropy constant $K_1$, MJ/m$^3$ |
|---|---|---|---|---|
| AlNiCo [1] | 10 – 40 | 1043 | 1,4 | 0,5 |
| FeN [1] | 160 | 810 | 1,92 | 1 |
| MnAl [1] | 112 | 650 | 0,6 | 1,7 |
| MnBi [1] | 106 | 633 | 0,58 | 0,9 |
| MnGa [1] | 69 | 770 | 0,47 | 2,35 |
| FePt [2] | 172 | 750 | 0,75 | 0,66 |
| CoCr [3] | 20 | 510 | 0,4 | 0,031 |
| HfCo$_5$ [4] | 278,4 | 751 | 1,18 | 1,67 |
| Zr$_2$Co$_{11}$ [4] | 188 | 783 | 0,97 | 1,35 |
| FeNi $L1_0$ [1] | 320 (theoretical) | 810 | 1,38 | 0,27 |

All the aforementioned methods [6-9] have been aimed at enhancement of atomic diffusion which controls formation of the $L1_0$ phase. To further increase the rate of $L1_0$ phase formation in FeNi alloys by 6–9 orders of magnitude, some authors [10] have recommended to focus the efforts on two main aspects: i) to increase the driving force for phase transformation, possibly through the addition of ternary alloying elements or additional stimuli that may enhance the stability of the $L1_0$ phase; ii) to increase the vacancy concentration in order to provide a large reservoir of grain boundaries and to enable atomic rearrangement, possibly through nanostructuring at low temperatures.

In this paper, we used the method of $L1_0$ FeNi synthesis by cyclic oxidation and reduction at 320 °C proposed by Lima et al [9, 11], which enormously enhances the local diffusion at the grain surfaces making the synthesis of Fe–Ni alloys at low temperatures possible. However, instead of nickel plated Fe particles of micrometer size, we subjected $Fe_{50}Ni_{50}$ nanopowder, produced by chemical co-precipitation, to the same heat treatments. It is expected that the usage of nanoparticles will further enhance the diffusion rate due to an increase of the surface contribution.

## 2. Experimental

### 2.1 Chemical co-precipitation

Precursors for chemical co-precipitation of FeNi were $Fe(NO_3)_3 \cdot 9H_2O$ and $Ni(NO_3)_2 \cdot 6H_2O$. Concentration of distilled water solutions were 10 %. Both solutions of the salts were pumped into a reactor, where they were mixed with 10 % NaOH solution in a ratio which allowed one to maintain the level of pH = 12. (Fe, Ni)(OH)$_2$ hydroxide, produced at the end of reaction, was rinsed in distilled water until pH reached 7 and after that it was centrifuged and dried in air at 50 °C for 72 hours. Obtained sample was then milled and placed in a tube furnace. Reduction of hydroxide was carried out in $H_2$ atmosphere at 320 °C for 4 hours.

Reactions occurring during process of production FeNi nanopowders by chemical precipitation method were as follow:

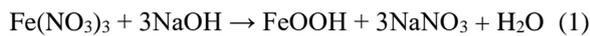
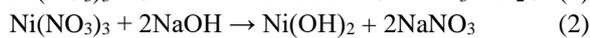
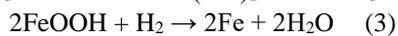
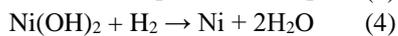

$$Fe(NO_3)_3 + 3NaOH \rightarrow FeOOH + 3NaNO_3 + H_2O \quad (1)$$
$$Ni(NO_3)_3 + 2NaOH \rightarrow Ni(OH)_2 + 2NaNO_3 \quad (2)$$
$$2FeOOH + H_2 \rightarrow 2Fe + 2H_2O \quad (3)$$
$$Ni(OH)_2 + H_2 \rightarrow Ni + 2H_2O \quad (4)$$

The first two reactions occured during process of precipitation in the reactor. The last two reactions took place in the tube furnace during the reduction in $H_2$ atmosphere.

### 2.2 Cyclic oxidation and reduction process

Obtained FeNi powder was subjected to heat treatment, which included oxidation for 20 min under $N_2/O_2$ atmosphere with a ratio of 95/5 and subsequent reduction under $H_2$ atmosphere during 40 min. We performed 5 and 10 cycles of the heat treatment.

### 2.3 Characterization

Morphology of the samples was investigated using scanning electron microscope (SEM) TESCAN Vega 3 SB. Elemental composition of the forming phases was identified by using Energy dispersive X-ray spectroscopy (EDX). Crystal structure and phase composition were studied by X-ray diffraction method (XRD) on Difrei 401 under Cr radiation. Differential scanning calorimetric analysis (DSC) was performed using NETZSCH STA 449F1 Jupiter calorimeter with a heating/cooling rate 10 °C/min in Ar atmosphere. Magnetic properties were investigated using vibrating sample magnetometer (VSM).

## 3 Results and discussion

Since the validation of the FeNi $L1_0$ phase formation is a challenging task, we used several experimental techniques including XRD, EDX, DSC and magnetic measurements to verify the possible presence of the desired phase in our samples.

SEM images of the FeNi powder sample just after chemical precipitation together with the samples subjected to 5 and 10 cycles of oxidation/reduction treatments are presented in Figure 1. A comparative analysis of morphology of the samples showed that processes of agglomeration and coalescence occur under the cyclic treatment. To determine average size of FeNi nanoparticles, corresponding calculations were performed on SEM images for at least 1000 nanoparticles by using common statistical methods and taking into account possible agglomeration of the particles. It was found that the average particles size was not significantly changed and for all the samples it was equal to 100 nm. In parallel, the average size of the FeNi nanoparticles was estimated from XRD data by means of the Debye-Scherrer equation:

$$L_C = \frac{k\lambda}{B\cos(\theta)},$$

where $L_C$ - crystallite size of coherent scattering (Å), $k$=0.89 is a dimensionless shape factor; $B$ is the line broadening at half of the maximum intensity (rad); $\lambda$ is wave length of the x-ray radiation (equal to 2,28975 Å for Cr Kα), θ is diffraction angle (rad). The accuracy of particle sizes calculations by the Debye-Scherrer formula is ± 5%. The value of coherent scattering regions of all the samples was 27 nm. The difference in particle size determination by means of two techniques is explained by the polycrystalline structure of the particles. As a consequence, each particle has several coherent scattering regions.

The chemical composition by means of EDX analysis of all samples was clarified to be 50:50. Figure 2 shows XRD patterns for the whole set of samples before and after cyclic oxidation and reduction treatments. As shown on the plots, all the samples consist of single fcc FeNi phase with the lattice parameter $a$=3.576 Å. EDX and XRD analyses revealed that there were no essential changes in the phase composition of the samples during heat treatments.

It should be mentioned that, according to the literature data [12], the intensity of the strongest (001) superstructure diffraction peak from a nontextured ordered tetrataenite phase would be only 0.3 % compared to the (111) diffraction peak [12]. Hence, since diffraction peaks of the FeNi $L1_0$ phase have very low intensity (fct tetrataenite forms superlattice), it is impossible to validate the presence of tetrataenite in the samples by means of XRD analysis.

In turn, recently Bordeaux et al. have reported [10] that a small endothermic peak (enthalpy is 4,4 kJ/mole) attributable to chemical disordering process of the tetrataenite phase is present in temperature range from 550 to 600 °C on DSC curves taken for the slices of meteorite that consisted of almost pure (approximately 92 %) FeNi $L1_0$ phase . Thus, DSC analysis can be very useful methods to confirm the presence of FeNi $L1_0$ phase in the powders.

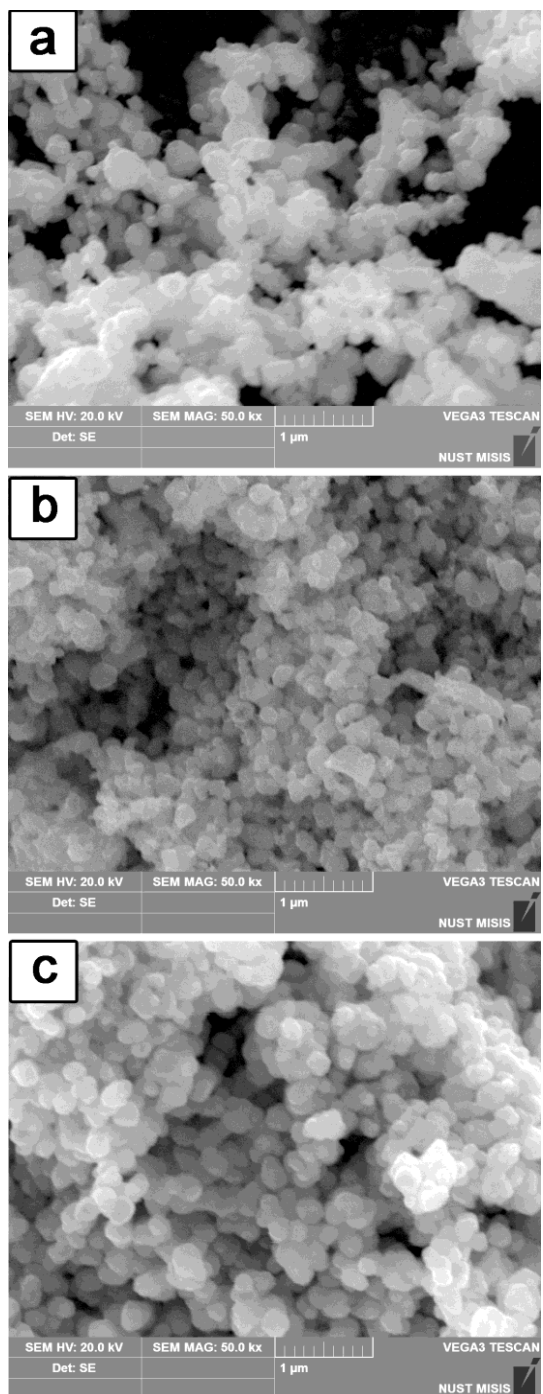

**Fig. 1**. SEM images of FeNi nanopowders: a – initial sample, b – after 5 cycles and c – after 10 cycles of the treatments.

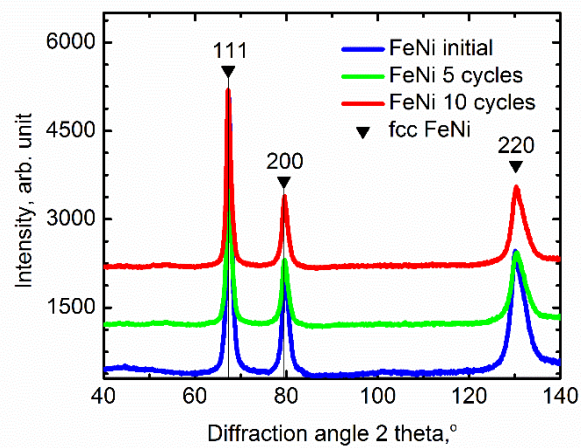

**Fig. 2.** XRD patterns of initial FeNi samples and the samples subjected to 5 and 10 cycles of treatments.

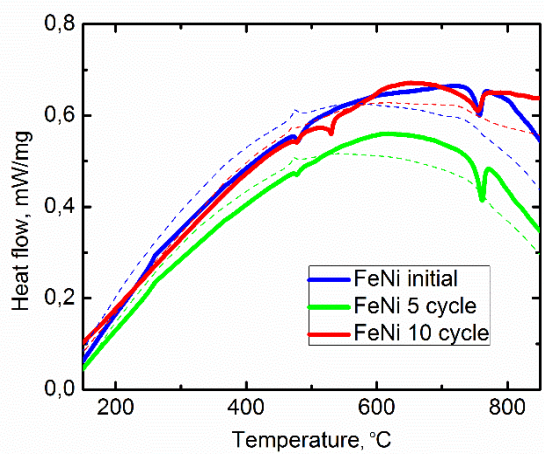

**Fig. 3.** DSC curves of initial FeNi samples and the samples subjected to 5 and 10 cycles of treatments upon heating (solid lines) and cooling (dashed lines).

DSC curves measured upon heating (solid line) and cooling (dashed line) of the samples are presented in Figure 3. The heat flow vs temperature dependence for all the sample shows identical behavior with anomalies at 480 °C and 730 °C. The endothermic peak at 730 °C corresponds to formation of nickel ferrite from Ni and Fe oxides on the surface of the particles [13, 14], which in turn were synthesized during chemical precipitation due to nanopowders' pyrophorosity. The absence of the oxide and nickel ferrite phases on the XRD patterns (see Figure 2) is attributed to inability of this method to register low content phases (less than 5 %). The endothermic peak at 480 °C is associated with Curie temperature of the fcc FeNi [15]. The small misfit with the literature data is explained by the fact that nanoscaled magnetic materials reveal a lower value of $T_C$ due to surface effect [16, 17].

It is worth mentioning that the sample subjected to 10 cycles of oxidation/reduction treatment revealed additional endothermic peak at 530 °C. Temperature of the peak lies exactly in the range corresponding to disordering process of the $L1_0$ FeNi phase. DSC curves measured under cooling protocol showed all peaks except the peak at 530 °C that is explained by an irreversible character of the chemical disordering process of the tetrataenite phase in the sample. To be convinced of the correctness of the conclusions based on the DSC analysis data and for further confirmation of the tetrataenite phase formation, the investigation of magnetic properties were carried out.

According to the literature data, FeNi $L1_0$ phase reveals high values of magnetocrystalline anisotropy field $H_A$=20 kOe [18]. In order to evaluate anisotropy field, each powder samples was textured in an external magnetic field and compacted with epoxy to the cylindrical shape with the diameter to height ratio of 1. This corresponds to demagnetizing factor N=1/4 along the texture direction. Figure 4 shows field dependences of magnetization after form factor correction for the FeNi samples before and after heat treatment. Measurements along easy axis (EA) correspond to those performed in the external magnetic field applied along the texture direction. As shown on the graphs, only the sample subjected to 10 cycles of oxidation/reduction treatments revealed a coercitive field of $H_C$ = 500 Oe, which is indicative of the presence of the $L1_0$ phase in the specimen. Two other samples showed the typical soft magnetic behavior of magnetization with the coercivity of $H_C$=50 Oe. The presence of the latter is explained by small FeNi particle size. However, the magnetization curves along hard axis for the sample subjected to 10 cycles of oxidation/reduction treatments did not show sizeable anisotropy. This is explained by the challenges with assembling nanoparticles in an aligned structure. This is a common problem with the measurements of magnetic properties of nanoscale particles, caused by the trend of nanoparticles to aggregate in clusters. Magnetic moment orientation in that clusters sums up from average magnetic moment orientations of the constituent particles.

To solve this problem and to evaluate the anisotropy field it was proposed to use Singular Point Detection (SPD) technique [19, 20]. It is very powerful technique for the study of anisotropy in polycrystalline samples. The authors [19, 20] proceed from the assumption that when we apply a magnetic field to a polycrystalline material we obtain, in general, a smooth M(H) curve which, at first sight, seems to give no indication of the singularity at $H = H_A$. However, averaging over all the crystal orientations does not cancel the singularity coming from the contribution of the crystallites oriented in such a way that their hard directions are nearly parallel to H. The singularity can be detected by observation of the successive derivatives $d^2M/dH^2$. Thus, in this method $H_A$ can be estimated from the graphs of the second derivative of magnetization curves measured along easy axis. Calculated SPD curves for the whole set of the samples are presented in Figure 5. As a source for the SPD curves we used initial magnetizing curves. In all the graphs, the first main peak is associated with reaching the saturation magnetization of the majority of the FeNi particles. It should be mentioned that only in the sample subjected to 10 cycles of oxidation/reduction treatments the second peak attributed to FeNi $L1_0$ grains reversal is observed. An anisotropy field $H_A$ for this sample can be calculated as a difference between two peaks observed on the SPD curve. The estimated value of $H_A$ = 18 kOe in the FeNi sample after 10 cycles of thermal treatment is in a good agreement with the literature data.

Thus, by means of DSC method together with magnetic measurements we confirmed the possibility of the tetrataenite phase formation in nanoparticle samples subjected to cyclic oxidation and reduction treatments.

The quantitative analysis of the DSC data by means of calculating the enthalpy of the reaction, which in turn is equal to the peak area, allowed us to estimate the volume fraction of the desired phase. According to the experimental data (see Figure 3), enthalpy of the endothermic reaction at 530 °C $\Delta H$=0,44 kJ/mole which is of order of magnitude smaller as compared with that obtained for meteorite consisting almost pure FeNi (~ 92 %) $L1_0$ phase. Thus, mass fraction of the tetrataenite phase in our sample was equal to ~10 wt.%.

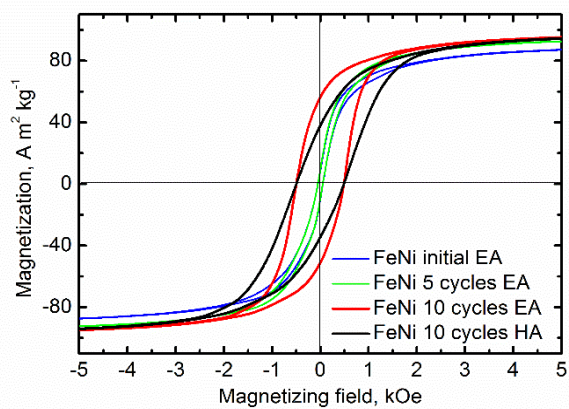

**Fig. 4.** Magnetization curves of initial FeNi samples and the samples subjected to 5 and 10 cycles of treatments.

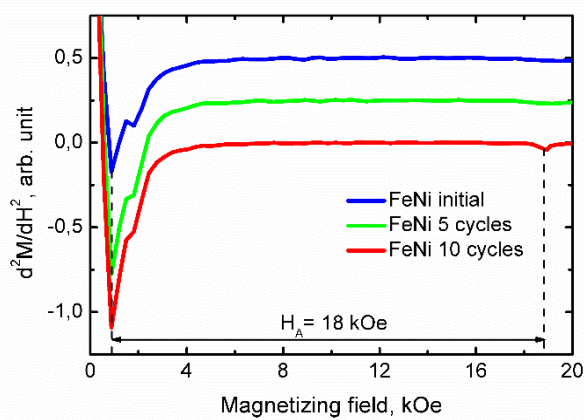

**Fig. 5.** SPD curves of initial FeNi samples and the samples subjected to 5 and 10 cycles of treatments.

**4. Conclusions**

By using cyclic oxidation and reduction at 320 °C we synthesized FeNi L10 (tetrataenite) phase in chemically co-precipitated FeNi nanopowder. Successful production of tetrataenite phase in the samples was confirmed by DSC analysis and magnetic measurements. In the sample subjected to 10 cycles of oxidation / reduction treatments the DSC curves showed endothermic peak at temperature of 530 °C which correspond to chemical disorder transformation of tetrataenite. Evaluated by means of SPD method value of HA = 18 kOe is close to experimental data for FeNi L10 phase. To enhance the anisotropy field, which in turn is a key factor for permanent magnets application, and to produce textured material it is proposed to modify shape of the particles and bring it to the anisotropic form, for instance nanorods or nanotubes. Developed in our work approach enhances the diffusion rate in solid state and can be used for stabilization of 'new' compounds in a low-temperature region (below 320 °C) of phase diagrams of materials with a high melting temperature (above 1500 °C). Some of these low temperature phases could become new substitution phases with extreme magnetic characteristics and low production costs.

**Acknowledgements**

This work was supported by Russian Science Foundation Grant No. 15-12-10008. D. Karpenkov gratefully

acknowledges the financial support of the Ministry of Education and Science of the Russian Federation in the framework of Increase Competitiveness Program of NUST "MISiS" (No K4-2015-013). A. Karpenkov gratefully acknowledges the financial support of the State task "Ensuring the Performance of the Scientific Research".
## References

[1] J.M.D. Coey, Permanent magnets: Plugging the gap, Scr. Mater. 67(6) (2012) 524–529.

[2] J. Wang, B. Yang, W. Pei, et al., Structural and magnetic properties of L1$_0$-FePt/Fe exchange coupled nano-composite thin films with high energy product, J. Magn. Magn. Mater. 345 (2013) 165–170.

[3] V.G. Harris, Y. Chen, A. Yang, et al., High coercivity cobalt carbide nanoparticles processed via polyol reaction: a new permanent magnet material, J. Phys. D: Appl. Phys. 43 (16) (2010) 165003.

[4] B. Balamurugan, B. Das, W.Y. Zhang, et al. Hf–Co and Zr–Co alloys for rare-earth-free permanent magnets, J. Phys. Condens. Matter. 26 (6) (2014) 064204.

[5] L. Néel, J. Pauleve, R. Pauthenet, et al., Magnetic Properties of an Iron–Nickel Single Crystal Ordered by Neutron Bombardment, J. Appl. Phys. 35 (1964) 873–876.

[6] T. Shima, M. Okamura, S. Mitani, K. Takanashi, Structure and magnetic properties for L1$_0$-ordered FeNi films prepared by alternate monatomic layer deposition, J. Magn. Magn. Mater. 310 (2007) 2213–2214.

[7] S. Lee, K. Edalati, H. Iwaoka et al., Formation of FeNi with L10-ordered structure using high-pressure torsion, Philos. Mag. Lett. 94(10) (2014) 639–646.

[8] A. Makino, P. Sharma, K. Sato, et al., Artificially produced rare-earth free cosmic magnet, Sci. Rep. 5 (2015) 16627.

[9] E. Lima, V. Drago, P. Fichtner, et al., Tetrataenite and other Fe–Ni equilibrium phases produced by reduction of nanocrystalline NiFe$_2$O$_4$, Solid State Comm. 128 (9–10) (2003) 345–350.

[10] N. Bordeaux, A.M. Montes-Arango, J. Liu, et al., Thermodynamic and kinetic parameters of the chemical order–disorder transformation in L1$_0$ FeNi (tetrataenite), Acta Mater. 103 (2016) 608–615.

[11] E. Lima, V. Drago, A New Process to Produce Ordered Fe$_{50}$Ni$_{50}$ Tetrataenite, Phys. Status Solidi. 128 (1) (2001) 119–124.

[12] A.M. Montes-Arango, L.G. Marshall, A.D. Fortes et al., Discovery of process-induced tetragonality in equiatomic ferromagnetic FeNi, Acta Mater. 116 (2016) 263–269.

[13] J. Beretka, A.J. Marriage, Determination of the Initial Formation Temperature of Nickel Ferrite Spinel, Nature 203 (1964) 515.

[14] J. Cassedanne, Study of the binary diagrams α-Fe$_2$O$_3$-NiO and La$_2$O$_3$-NiO and the ternary diagram α-Fe$_2$O$_3$-NiO-La$_2$O$_3$, Anais Acad. Brasil. Cien. 36 (1) (1964) 13.

[15] W. Xiong, H. Zhang, L. Vitos, et al., Magnetic phase diagram of the Fe–Ni system, Acta Materialia 59 (2011) 521–530.

[16] X. Batlle, A. Labarta, Finite-size effects in fine particles: magnetic and transport properties, J. Phys D.: Appl. Phys. 35 (6) (2002) R15.

[17] R.H. Kodama, Magnetic nanoparticles, J. Magn. Magn. Mater. 200 (1–3) (1999) 359–372.

[18] L.H. Lewis, A. Mubarok, E. Poirier, et al., Inspired by nature: investigating tetrataenite for permanent magnet applications, J. Phys.: Condens. Matter. 26 (6) (2014) 064213.

[19] G. Asti, S. Rinaldi, Nonanaliticity of the Magnetization Curve: Application to the Measurement of Anisotropy in Polycrystalline Samples, Phys. Rev. Lett. 28 (24) (1972) 1584–1586.

[20] F. Bolzoni, R. Cabassi, Review of singular point detection techniques, Physica B 346–347 (2004) 524–527.
5